\documentstyle[12pt]{article}

\catcode`@=11
\def\secteqno{\@addtoreset{equation}{section}%
\def\theequation{\thesection.\arabic{equation}}}
\catcode`@=12
\secteqno
\newcommand{\be}{\begin{equation}}
\newcommand{\ee}{\end{equation}}
\newcommand{\bea}{\begin{eqnarray}}
\newcommand{\eea}{\end{eqnarray}}
\newcommand{\bref}[1]{(\ref{#1})}
\newcommand{\nn}{\nonumber}

\newcommand{\mapright}[1]{\smash{[mathop{\hbox to 1cm{\rightarrowfill}}
\limits^{#1}}}

\newcommand{\N}{\mathcal{N}}

\begin{document}

\begin{flushright}
\parbox{4.2cm}
{2012,~August  14\\
KEK-TH-1562 \hfill \\
}
\end{flushright}

\vspace*{1.1cm}

\begin{center}
 \Large\bf  SL(5) duality from canonical M2-brane 
\end{center}
\vspace*{1.5cm}
\centerline{\large Machiko Hatsuda$^{\dagger\ast a}$ and Kiyoshi Kamimura$^{\star b}$}
\begin{center}
$^{\dagger}$\emph{Physics Department, Juntendo University, 270-1695, Japan}
\\
$^{\ast}$\emph{KEK Theory Center, High Energy Accelerator Research 
Organization,\\
Tsukuba, Ibaraki 305-0801, Japan} 
\\
$^{\star}$\emph{Department of Physics, Toho University, Funabashi, 274-8510, Japan
}
\vspace*{0.5cm}
\\
$^{a}$mhatsuda@post.kek.jp
~~;~~
 $^{b}$kamimura@ph.sci.toho-u.ac.jp
\end{center}

\vspace*{1cm}

\centerline{\bf Abstract}
  
\vspace*{0.5cm}
We show how
the SL(5) duality in M-theory is explained from a canonical analysis 
of  M2-brane mechanics.
Diffeomorphism constraints for a
 M2-brane coupled to supergravity background in $d=4$
are reformulated in a SL(5) covariant form,
 in which spatial diffeomorphism constraints 
are  recast into  a SL(5) vector
and the generalized metric in the Hamiltonian constraint
is quartic in the SL(5) generalized vielbein.
The Hamiltonian for a M2 brane has the SL(5) duality symmetry 
in a background dependent gauge. 
 \vfill 

\thispagestyle{empty}
\setcounter{page}{0}
\newpage 
\section{Introduction}

 In contrast to the pure gravity theory,
string gravity and membrane gravity theories
contain rich duality symmetries
governed by generalized geometry 
introduced  by  Hitchin \cite{Hitchin:2004ut}, 
Gualtieri \cite{Gualtieri} for string system and 
Hull   \cite{Hull:2007zu}  for M theory.
For these theories the general coordinate transformation is generalized
to the gauge transformation for both gravity field and
 gauge fields coupled to extended objects 
and it is  given by Courant bracket or 
C-bracket \cite{Siegel:1993xq,Siegel:1993th}.
Doubled formalism with manifest T-duality
 \cite{{Tsey},{Hull:2004in}} 
 and double field theory 
formulated by \cite{{Hull:2009mi}}
are complementary approaches to understand duality.
Further  studies appeared in  
 \cite{{Bonelli:2005ti},
 {Pacheco},{Coimbra:2011nw},{Hohm:2010pp}, {Hohm:2010xe},{Berman:2011cg}, {Jeon:2012kd},{Thompson:2011uw},{Andriot:2011uh},
{Hohm:2011nu}} and 
\cite{{Hassan:1999bv},
{Koerber:2005qi},{Albertsson:2008gq},
{Hohm:2011zr},
{Hatsuda:2012uk},
{Kikuchi:2012za}} on D-branes.

For a bosonic string theory 
the gravitational field $G_{mn}$ and
the rank two tensor $B_{mn}$ field are 
mixed by the T-duality symmetry,
where the relation between
 T-duality in the first quantized level
and the one in the second quantized level 
is well known.
On the other hand  U-duality in the first quantized level
and the one in  M-theory
have not been fully understood so far. 
The U-duality,  as the symmetry of solitonic charges 
associated with extended  objects 
\cite{Hull:1994ys},
  is the hidden symmetry in 
the 11-dimensional supergravity theory
\cite{Cremmer:1979up}.
The relation between the U-duality 
and a membrane duality was shown 
by  Duff and Lu  \cite{Duff:1990hn}
 using with the Gaillard-Zumino's (GZ) dual
 formulation 
\cite{Gaillard:1981rj}.
Recently the supergravity is reformulated in a manifest  duality covariant way 
by Berman and Perry \cite{Berman:2011cg},
relating to the membrane duality which is also 
 treated  in the GZ  dual  formulation.

 In this paper we  examine the membrane duality 
in  canonical language 
instead of the GZ dual formulation.
We focus on the SL(5) duality for the case of $d=4$.
In reference \cite{Hatsuda:2012uk} 
one of the present author
has made a canonical analysis of D-brane mechanics
 to obtain Courant brackets 
by using which the
general coordinate  transformation and the gauge transformation 
for D-branes are derived.
In this paper we apply the above analysis to a  M2 brane,
and we clarify the relation between SL(5) duality 
in M2 mechanics  and the one in the supergravity theory.
  
One of the typical features of actions of extended objects
 is the diffeomorphism invariance.
For a string case the $\sigma$-diffeomorphism constraint,
which corresponds to 
$\triangle=$$\frac{\partial}{\partial x^m}\frac{\partial }{\partial \tilde{x}_m}=0$ constraint for example
in \cite{Hull:2009mi}, plays an essential role.
Diffeomorphism constraints for a bosonic string
are written in terms of  $Z_M=(p_m,\partial_\sigma x^m)$ 
which is the basis of 
the generalized geometry,
where $x^m$ and $p_m$ are string coordinate and its conjugate momentum
with $m=1,\cdots,d$.
The $\sigma$-diffeomorphism constraint
${\cal H}_\sigma$ and the $\tau$-diffeomorphism constraint, 
Hamiltonian  ${\cal H}_\perp$,  for a string
 are written as
\bea
\left\{
{\renewcommand{\arraystretch}{1.4}
\begin{array}{lcl}
{\cal H}_\sigma=\frac{1}{2}Z_M~\rho^{MN}~Z_N=0&,&
\rho^{MN}=
\left(\begin{array}{cc}
0&\delta_m^n\\\delta_n^m&0
\end{array}\right)\\
{\cal H}_\perp =\frac{1}{2}Z_M~{\cal M}^{MN}~Z_N=0&,&
{\cal M}^{MN}=\left(
\begin{array}{cc}
G^{mn}&-G^{mq}B_{qn}\\
B_{mp}G^{pn}~~~&G_{mn}-B_{mp}G^{pq}B_{qn}
\end{array}
\right)
\end{array}}\right.\label{strho}
\eea
where ${\cal M}^{MN}$ is the generalized metric as a function of
 $G_{mn}$ and $B_{mn}$.
The $\sigma$ -diffeomorphism constraint ${\cal H}_\sigma$
has O($d,d$) invariance. 
 Under the O($d,d$)$\ni g$ transformation, $Z\to gZ$, 
 the Hamiltonian is covariant
since $G$ and $B$ are coset parameters of O($d,d$)/O($d$)$\times$O($d$):
${\cal M}(G,B)\to {\cal M}'=g^T{\cal M}g={\cal M}(G',B')$
and so ${\cal H}_\perp(G,B)\to {\cal H}_\perp(G',B')$. 

On the other hand the fundamental basis for a M2-brane
are $
Z_M=(p_m,~\frac{1}{2}\epsilon^{ij}\partial_i x^m\partial_i x^n)$,
$i=(1,2)$.
Invariant symmetry of  $\sigma^i$ -diffeomorphism constraints,
${\cal H}_i=p_m\partial_i x^m=0$, is not apparent
in this form.
 Multiplying $\epsilon^{ij}\partial_j x^p $    on ${\cal H}_i$ 
 makes ${\cal H}^p$ to be bilinear in   $Z_m$,
 although the metric $\tilde{\rho}^{MN;p}$  is not manifestly 
 duality symmetry invariant  
  as seen in \cite{Hatsuda:2012uk}.
In this paper we  pursue the diffeomorphism  constraints
in focusing on $d=4$ case leading to the SL(5) duality.

The organization of this paper is the following:
In section 2  a canonical analysis of a M2-brane
is presented in  the basis of the generalized geometry.
In section 3 we show that 
a Courant bracket for a M2 brane  is calculated and 
 generalized gauge transformation 
of  $G_{mn}$ 
and $C_{mnl}$ are derived  by
 using the Courant bracket. 
In section 4 we examine the invariance of the diffeomorphism
constraints for $d=4$ case.
We show that $\sigma^i$-diffeomorphism constraints are recast into a SL(5) vector  
  in terms of the basis in a form of SL(5) rank two tensor,  
resulting SL(5) invariance on the constrained surface.
The Hamiltonian constraint is also rewritten in terms of 
the tensor basis, 
and the generalized metric is rewritten as
a rank four tensor, antisymmetric in a pair of two indices and symmetric in
pairs, parameterized by  $G_{mn}$ and $C_{mnl}$. 
We show that the generalized metric is written in a quartic form of the generalized vielbein
given in $T^4$ compactified supergravity theory
\cite{{Cremmer:1979up},
{Sezgin:1982gi},{Tanii:1984zk}}. 
Then reformulated diffeomorphism constraints are manifestly
SL(5) covariant in a gauge which depends on the background.
It leads to SL(5) duality symmetry transformation 
for supergravity fields.

\par\vskip 6mm
\vskip 6mm
\section{Hamiltonian for a M2-brane }

We begin with an action for a M2-brane 
\bea
I&=&\displaystyle\int_M d^3\sigma~({\cal L}_{0}+{\cal L}_{WZ})
\nn\\
{\cal L}_{0}&=&-T_{M2}
\sqrt{-h}~~,~~
h=\det h{}_{\mu\nu}~~,~~
 h{}_{\mu\nu}=\partial_\mu x^m \partial_\nu x^n G_{mn}
\label{action}\\
{\cal L}_{WZ}&=&
 \displaystyle\frac{1}{3!}T_{M2}\epsilon^{\mu\nu\rho}
~\partial_{\mu}x^m\partial_\nu x^n \partial_{\rho}x^l C_{mnl}
~~~\nn
\eea
where   $G_{mn}(x)$ is the background metric and
$C_{mnl}(x)$ is a rank three anti-symmetric gauge field.
We focus on the bosonic part only throughout this paper.
Target space indices, the 3-dimensional world volume indices
and  spatial world volume indices
are denoted by $m,~n,\cdots=0,\cdots,d-1$;
 $\mu, ~\nu,~\cdots =0,~1,~2$ 
and $i,~j,~\cdots=1,~2$ respectively.   
The canonical momenta are defined as 
\bea
p_m
&=&{T_{M2}}\left(
-\sqrt{-h}h_{}^{0\mu}G_{mn}\partial_\mu x^n
+{\frac12}\epsilon^{0ij}
\partial_i x^n\partial_j x^l
 C_{mnl}\right)~~~.\nn
\eea
The  Hamiltonian constraint and the $\sigma^i$-diffeomorphism constraints are
\bea
&&\left\{
{\renewcommand{\arraystretch}{1.4}
\begin{array}{ccl}
{\cal H}_\perp &=&\displaystyle\frac{1}{2T_{M2}}
\left(
\tilde{p}_m G^{mn}\tilde{p}_n
+T_{M2}{}^2 \det h{}_{ij}
\right)~=~0\\
{\cal H}_i&=&\partial_i  x^m\tilde{p}_m
~=~
\partial_i  x^m{p}_m~=~0
\end{array}}\right.\label{HamM2}
\eea
with 
\bea
\tilde{p}_m&\equiv&p_m
- \frac{T_{M2}}{2}\epsilon^{0ij}
\partial_i x^n\partial_j x^l
C_{mnl}
\nn\\
&=&-T_{M2}\sqrt{-h}
h{}^{\mu 0}G_{mn}\partial_\mu x^n~~\nn~,
\eea
similar to the IIA D2 brane case
\cite{Hatsuda:1998by}.
The determinant term can be rewritten as
\bea
\det h_{ij}&=&\frac{1}{2}
(\epsilon^{ij}\partial_i x^m \partial_jx^n )
G_{mm'}G_{nn'}
(\epsilon^{i'j'}\partial_{i'} x^{m'} \partial_{j'}x^{n'}) 
~~~.\nn
\eea
For simplicity we take a unit $T_{M2}=1$
throughout the rest of this paper.
Then the Hamiltonian constraint for the M2-brane in curved background
is given by the $Z_M$ basis and the generalized metric as
\bea
{\cal H}_\perp&=&\frac{1}{2}
Z_M{}~{\cal M}^{MN}~Z_N\nn\\
&&Z_N~=~\left(
{\renewcommand{\arraystretch}{1.2}
\begin{array}{c}
p_n\\
\frac{1}{2}\epsilon^{ij}\partial_i x^n \partial_jx^{n'}
\end{array}}
\right)\nn\\
&&{\cal M}^{MN}~=~
~\left(
{\renewcommand{\arraystretch}{1.2}
\begin{array}{cc}
G^{mn}&-G^{mk}C_{knn'}\\
-C_{mm'l}G^{ln}~~~&
G_{[m|n}G_{|m']n'}
+C_{mm'l}G^{lk}C_{knn'}
\end{array}}
\right)~~~.\label{MM2}
\eea
It is rewritten as
\bea
{\cal M}^{MN}&=&({\cal N}^{T})^{M}{}_{L}{\cal M}_0{}^{LK}
{\cal N}_{K}{}^{N} \nn\\
{\cal M}_0^{LK}&=&
\left(
\begin{array}{cc}
G^{lk}&0\\
0&G_{[l|k}G_{|l']k'}
\end{array}
\right)~~,~~{\cal N}_{K}{}^{N}~=~
\left(
\begin{array}{cc}
\delta_k{}^n&-C_{knn'}\\
0&\delta_n^k \delta^{k'}_{n'}
\end{array}
\right)
\nn~~~.
\eea
It is further rewritten as
\bea
{\cal M}_0^{MN}&=&(\mu^T)^M{}_A\eta^{AB}\mu_B{}^N~~,~~
\eta^{AB}~=~
{\renewcommand{\arraystretch}{1.2}
\left(
\begin{array}{cc}
\delta^{ab}&0\\0&\delta_{[a|b}\delta_{|a']b'}
\end{array}
\right)~~,~~
\mu_{A}{}^K=
\left(
\begin{array}{cc}
e_a{}^k&0\\0&\frac{1}{2}e_{[k}{}^ae_{k']}{}^{a'}
\end{array}
\right)}\nn\\
{\cal M}^{MN}&=&(\nu^T)^M{}_A\eta^{AB}\nu_B{}^N~~,~~
\nu_A{}^N=
\mu_A{}^K
{\cal N}_K{}^N~=~
{\renewcommand{\arraystretch}{1.2}\left(
\begin{array}{cc}
e_a{}^n&-e_a{}^kC_{knn'}\\0&\frac{1}{2}e_{[n}{}^ae_{n']}{}^{a'}
\end{array}
\right)}~~~,\label{munu}
\eea
with $G_{mn}=e_m{}^ae_n{}^b\eta_{ab}$.
The  diffeomorphism constraints ${\cal H}_i (i=1,2)$
can be also written in terms of $Z_M$ basis by contracting with
$ \epsilon^{ij}\partial_jx^p$
\bea
{\cal H}^p &\equiv&\frac{1}{2}\epsilon^{ij}\partial_i x^p {\cal H}_j~=~
\frac{1}{2}Z_M\tilde{\rho}^{MN:p}Z_N~=~
0 ~~,~~
\tilde{\rho}^{MN;p}=
\left(\begin{array}{cc}
0&\delta^p_{[n}\delta^m_{l]}\\
\delta^p_{[m}\delta^n_{l]}&0
\end{array}\right)\label{Hj}
\eea
where ${\cal H}^p  (p=0,...,d)$ are reducible.

Now $Z_M$ is a fundamental basis of the canonical analysis.
The $Z_M$ algebra is given by
\bea
&
\left\{Z_M(\sigma),Z_N(\sigma')\right\}=i\rho_{MN}^i
\partial_i \delta(\sigma-\sigma')~~,~~
\rho_{MN}^i=
\left(
{\renewcommand{\arraystretch}{1.2}
\begin{array}{cc}
0&
\frac{1}{2}\epsilon^{ij}\partial_j x^{[n}\delta_m^{l]}\\
\frac{1}{2}\epsilon^{ij}\partial_jx^{[m}\delta_n^{l]}&0
\end{array}}
\right)&\label{M2rho}
\eea
where $\rho_{MN}^i$ is divergenceless,  $\partial_i\rho_{MN}^i=0.$
The metric $\rho_{MN}^i$ in \bref{M2rho} reduces to the metric $\tilde{\rho}^{MN;p}$
in \bref{Hj}   for the ground state in static gauge,
$\partial_jx^p=\delta_j^p$.  

 \par\vskip 6mm
\vskip 6mm
\section{Courant bracket for M2 brane}

In this section we will write down a Courant bracket
for a M2 brane  explicitly
in the notation of \cite{Hatsuda:2012uk}.
Let us consider a space generated by the algebra in 
\bref{M2rho}.
A vector in the space is given by
 \bea
& 
\hat{\Lambda}~=~\Lambda^M Z_M~=~
\lambda+\lambda^{[2]}
~\in~
T  \oplus \Lambda^{2}T^\ast 
&\nn\\
&\lambda=\Lambda^m p_m~~,~~
\lambda^{[2]}=\frac{1}{2}
\Lambda_{mn}~\frac{1}{2}\epsilon^{ij}\partial_i x^m\partial_j x^n~~~.&
\eea
The canonical commutator between two vectors $\hat{\Lambda}_1(\sigma)$ 
and $\hat{\Lambda}_2(\sigma')$ is given by
\bea
\left\{\hat{\Lambda}_1(\sigma),\hat{\Lambda}_2(\sigma')\right\}
=-i\hat{\Lambda}_{12}(\sigma)
\delta(\sigma-\sigma')
+i\left(
\displaystyle\frac{1-K}{2}\Psi_{(12)}^i(\sigma)
-\displaystyle\frac{1+K}{2}\Psi_{(12)}^i(\sigma')
\right)\partial_i\delta(\sigma-\sigma')\nn\\
\label{L1L2}
\eea
with
\bea
\hat{\Lambda}_{12}&=&
\Lambda_{[1}{}^l\partial_l\Lambda_{2]}{}^m~p_m\nn\\
&&+\frac{1}{2}\left({\cal L}_{\lambda_{[1}}\lambda_{2]}^{[2]}
-\partial_{[m|}(\Lambda_{[1}{}^l\Lambda_{2]l|n]})
+K\partial_{[m|}(\Lambda_{(1}{}^l\Lambda_{2)l|n]})
\right)\frac{1}{2}\epsilon^{ij}\partial_ix^m\partial_jx^n\nn\\
\Psi_{(12)}^i&=&\Lambda_{(1}{}^m\Lambda_{2)mn}\frac{1}{2}\epsilon^{ij}
\partial_jx^n~=~\frac{1}{2}\Lambda_1{}^M\Lambda_2{}^N \rho_{MN}^i~~~\label{phiphi}\\
{\cal L}_{\lambda_{[1}}\lambda_{2]}^{[2]}&=&
\frac{1}{2}\left(\Lambda_{[1}^l\partial_l\Lambda_{2]mn}
 +\partial_{[m|}\Lambda_{[1}^{l}\Lambda_{2]l|n]}
\right)\frac{1}{2}\epsilon^{ij}\partial_i
x^m\partial_jx^n\nn~~~.
\eea
 $K$ is an arbitrary constant
reflected by an ambiguity of $\partial_i\delta(\sigma-\sigma')$
as shown in \cite{Hatsuda:2012uk}.
Courant bracket for a M2-brane
is given as $\hat{\Lambda}_{12}$ part in \bref{L1L2};
\bea
[\hat{\Lambda}_1,\hat{\Lambda}_2]_{M2}
&=&[\lambda_1,\lambda_2]
+{\cal L}_{\lambda_{[1}}\lambda_{2]}^{[2]}
-d\left(
\iota_{\lambda_{[1}} \lambda_{2]}^{[2]}\right)
\label{M2}
\eea
\bea
&
\left\{
{\renewcommand{\arraystretch}{1.2}
\begin{array}{ccl}
[\lambda_1,\lambda_2]&=&
\Lambda_{[1}^n\partial_n\Lambda_{2]}{}^m p_m\\
{\cal L}_{\lambda_{[1}}\lambda_{2]}^{[2]}&=&
\frac{1}{2}\left(\Lambda_{[1}^l\partial_l\Lambda_{2]mn}
 +\partial_{[m|}\Lambda_{[1}^{l}\Lambda_{2]l|n]}
\right)\frac{1}{2}\epsilon^{ij}\partial_i
x^m\partial_jx^n\\
d(\iota_{\lambda_{[1}}\lambda^{[2]}_{2]})&=&\frac{1}{2}
\partial_{[m|}(\Lambda_{[1}^l\Lambda_{2]l|n]})~
\frac{1}{2}\epsilon^{ij}\partial_ix^m\partial_jx^n
\end{array}}\right.&\nn
\eea
for $K=0$,
and 
\bea
&[\hat{\Lambda}_1,\hat{\Lambda}_2]_{M2}
~=~[\lambda_1,\lambda_2]
+{\cal L}_{\lambda_{1}}\lambda_{2}^{[2]}
-\iota_{\lambda_{2}} d\lambda_{1}^{[2]}&
\label{M2half}\label{M2Courant}\\\nn
\\&
\left\{\begin{array}{ccl}
{\cal L}_{\lambda_{1}}\lambda_{2}^{[2]}&=&
\frac{1}{2}\left(\Lambda_{1}^l\partial_l\Lambda_{2mn}
 +\partial_{[m|}\Lambda_{1}^{l}\Lambda_{2l|n]}
\right)\frac{1}{2}\epsilon^{ij}\partial_i
x^m\partial_jx^n\\
\iota_{\lambda_{2}}d\lambda^{[2]}_{1}&=&
\Lambda_{2}^l\partial_{[l|}\Lambda_{1|mn]}^{[2]}
\frac{1}{2}\epsilon^{ij}\partial_ix^m
\partial_jx^n\end{array}\nn\right.&
\eea
for $K=1$.

Now let us calculate the generalized gauge transformation.
It is convenient to introduce
\bea
(\hat{C}_a)^M=\left(
\begin{array}{c}
e_a{}^m\\
e_a{}^lC_{mnl}
\end{array}\right)~~
,~~e_m{}^a e_a{}^n=\delta_m^n~~
\label{Chat}
\eea
where $a$ is the local SO($d$) index.
The gauge transformation rule is given by the Courant bracket 
in \bref{M2Courant} as
\bea
&&\delta_\xi(\hat{C}_a)^M=\left[
\hat{\xi},(\hat{C}_a)^M
\right]_{M2}~~,~~
\hat{\xi}=
{\renewcommand{\arraystretch}{1.2}
\left(
\begin{array}{c}
\xi^m
\\\xi_{mn}^{[2]}
\end{array}\right)}~~~.\label{gaugeC}
\eea
From the above transformation we obtain expected transformations 
of $G$ and $C$ by contracted the local SO($d$) indices,
\bea
&&~\left\{{\renewcommand{\arraystretch}{1.2}
\begin{array}{ccl}
\delta_\xi G_{mn}&=&
\xi^{l}\partial_{l} G_{mn}
+\partial_{(m|}\xi^{l}G_{l|n)}
 \\
\delta_\xi C_{mnl}&=&\xi^{p}\partial_{p} C_{mnl}
+\partial_{[m|}\xi^{p}C_{p|nl]}+\partial_{[m}\xi^{[2]}{}_{nl]}\end{array}}\right.
\label{M2}~~~.
\eea

It turns out that the gauge parameter 
$\xi^{[2]}{}_{mn}$ has a further gauge invariance,
namely gauge symmetry of gauge symmetry.
It is given by the invariance of
the Courant bracket.
We assume that  ${\Lambda}^M$'s are functions of only $x^m$.
Under  local transformations, 
$\delta \hat{\Lambda}_i$,~$i=1,2$,
invariance of the Courant bracket   
up to the total derivative  is given as 
\bea
&\delta \hat{\Lambda}_i~~\Rightarrow ~
\delta \displaystyle\int~d^2\sigma~
\left[\hat{\Lambda}_1,\hat{\Lambda}_2\right]_{M2}(\sigma)~=~0&~~~\label{gaugegauge}.
\eea
There is no further gauge symmetry of $\Lambda_i{}^m$ 
as seen from the
 coefficient of $p_m$ in \bref{gaugegauge}.
But there exists gauge symmetry of the parameter $\Lambda_{i;mn}^{[2]}$ 
as seen as below;
The coefficient of 
$\frac{1}{2}\epsilon^{ij}\partial_ix^m\partial_jx^n$ 
in \bref{gaugegauge}  for $K=1$  allows further gauge invariance as
\bea
\delta \Lambda_{i;mn}^{[2]}~=~\partial_{[m|}\zeta_{i|n]}
\Rightarrow ~
\delta \left({\cal L}_{\lambda_1}\lambda_2^{[2]}
-\iota{\lambda_2}d_{\lambda_2}d\lambda_1^{[2]}\right)&=&
\partial_m\left(
\Lambda_1{}^l\partial_{[l}\zeta_{2|n]}
\right)\frac{1}{2}\epsilon^{ij}\partial_i x^m
\partial_j x^n~\nn\\
&=&\partial_i\left(
\Lambda_1{}^l\partial_{[l}\zeta_{2|n]}
\frac{1}{2}\epsilon^{ij} x^m
\partial_j x^n\right)
~~.
\eea
Therefore gauge symmetry of the gauge parameter is given as
\bea
\delta \Lambda_i^M~=~
\left\{
\begin{array}{ccl}
\delta \Lambda_i{}^m
&=&0\\
\delta \Lambda_i{}^{[2]}_{mn}
&=&\partial_{[m|}\zeta_{i|n]}
\end{array}\right.
\eea
When the parameter is chosen as $\zeta_{i;n}=c_n\zeta_i$
 with a constant vector $c_m$,
the transformation becomes  
$\delta \Lambda_i=-c_{[m}\partial_{n]}\zeta_i$ 
as the survived component of $c_l\tilde{\rho}^{MN;l}\partial_N \zeta_i$.
The gauge symmetry of the 
gauge parameter $\hat{\xi}$ in \bref{M2}  is
$\delta \xi^m=0,~$
$\delta \xi_{mn}^{[2]}=
\partial_{[m}\zeta_{n]}$.

 \par\vskip 6mm
\vskip 6mm
\section{SL(5) duality}

Diffeomorphism constraint ${\cal H}_\sigma$
for a bosonic string theory
is invariant under T-duality symmetry.
However this is not apparent for a membrane theory.
The duality symmetry of the background field 
should be  a reflection of the duality symmetry of 
the world volume mechanics of $Z_M(\sigma)$.
So diffeomorphism  constriants for a  M2-brane should have
 U-duality symmetry. 

The pure gravity theory has global G=SL($d$) and local H=SO($d$) symmetries,
where the vielbein field is an element of the coset G/H;
$e_m{}^a~\to~g_m{}^ne_n{}^b h_{b}{}^{a}$ 
with G$\ni g$ and H$\ni h$.
Including to the shift of the dilatation field, GL(1), 
  it is extended to
the hidden symmetry which is
 larger  symmetry than  the manifest $d$-dimensional invariance
with subgroup SL($d$)$\times$GL(1) 
\cite{Cremmer:1979up}.
In this section we focus on $d=4$ case 
and  compare it  with the  known result 
of the SL(5) hidden symmetry of the supergravity theory.  
We analyze the symmetry of
$\sigma^i$  -diffeomorphism
constraints for a M2-brane at first.
Then we obtain the duality transformation
for $G_{mn}$ and  $C_{mnl}$ from the 
SL(5) transformation of the Hamiltonian for a M2 brane coupled to 
supergravity background,
which is  compared with the U-duality transformation 
of the  $T^4$ reduced supergravity theory.
At the same time we reformulate the diffeomorphism constraints 
and the Hamiltonian constraint in a  manifest SL(5) symmetric way.
\par

\subsection{SL(5) invariance of M2 diffeomorphism constraints}

Let us begin with
the $\sigma^i$ -diffeomorphism constraints 
in \bref{Hj}, ${\cal H}^p =Z_M ~\tilde{\rho}^{MN;p}~Z_N=0$,
 which would have 
 larger  symmetry than  the manifest $d$-dimensional invariance.
The number of basis $Z_M$ is 
$d+\frac{1}{2}d(d-1)=\frac{1}{2}d(d+1)$,
which is the number of the antisymmetric rank two tensor
in $(d+1)$-dimensions.
 A question is whether the  diffeomorphism constraint 
manifests the U-duality symmetry as was   the case of  
 string where the T-duality is realized as the O($d,d$) invariance
of the metric $\rho^{MN}$ in \bref{strho}.

In 4-dimensions the basis $Z_M$  is identified as
$10=4+6$-dimensional representation of SL(5).
It is rewritten as
a $5$-dimensional rank two anti-symmetric tensor 
as
\bea
Z_M=
{\renewcommand{\arraystretch}{1.2}
\left(
\begin{array}{c}
p_m\\\frac{1}{2}\epsilon^{ij}\partial_i x^{m}
\partial_j x^{n}
\end{array}
\right)}~
&\Rightarrow& ~
\check{Z}_{\hat{m}\hat{n}}=
{\renewcommand{\arraystretch}{1.2}
\left\{
\begin{array}{ccl}
\check{Z}_{\sharp n}
&=&p_{n}
\\
\check{Z}_{mn}&=&\frac{1}{2}
\epsilon_{m n pq}\epsilon^{ij}
\partial_i x^{p}
\partial_j x^{q}
\end{array}\right.}\label{tensor10}~~,~~\nn\\
&&~~~~~~~~~~~~~~~\hat{m}=(\sharp,m)~{\rm and}~ m=1,\cdots,4.
\label{eq4.1}
\eea
Similar representation have been used in
\cite{Cederwall:2007je}.
 Under infinitesimal 
SL(5) transformations 
\bea
A_{\hat{m}}{}^{\hat{n}}=\left(
\begin{array}{cc}
-\hat{  \alpha}&\gamma^n\\
\beta_m  & \alpha_m{}^n
\end{array}\right)~~~,~~\hat{\alpha}=\alpha_m{}^m~~~,
\eea 
covariant and contravariant vectors are transformed linearly as
$\delta u_{\hat{m}}~=~A_{\hat{m}}{}^{\hat{n}}u_{\hat{n}}$~,~
$\delta v^{\hat{m}}~=~-v^{\hat{n}}A_{\hat{m}}{}^{\hat{n}}$,
 so that $(v^{\hat{m}}u_{\hat{m}})$ is invariant.  
Thus the rank two tensor 
$\check{Z}_{\hat{m}\hat{n}}=(\check{Z}_{\sharp n},
~\check{Z}_{m n} )$
 in \bref{eq4.1} transforms as 
\bea
\delta
\check{Z}_{\hat{m}\hat{n}}&=&A_{[\hat{m}|}{}^{\hat{l}}
\check{Z}_{\hat{l}|\hat{n}]}\nn\\
&=&
\left\{
{\renewcommand{\arraystretch}{1.2}
\begin{array}{lcl}
\delta
\check{Z}_{\sharp n}&=&-\hat{\alpha}\check{Z}_{\sharp n}
+\alpha_n{}^l\check{Z}_{\sharp l}
+\gamma^l \check{Z}_{ln}\\
\delta 
\check{Z}_{m n}&=&\beta_{[m|}\check{Z}_{\sharp |n]}
+\alpha_{[m}{}^{{l}}\check{Z}_{l|n]}
\end{array}}\right.
~~~.\label{SL5abc}
\eea

In this basis the $\sigma^i$ -diffeomorphism constraint in \bref{Hj}
is written as
\bea
{\cal H}^p~=~\frac{1}{2}Z_M\tilde{\rho}^{MN;p}Z_N~=
 ~\epsilon^{mnlp} \check{Z}_{\sharp m}\check{Z}_{nl}=0~~~.
\label{Hdiffeoceck}
\eea
Under SL(5) in \bref{SL5abc} the diffeomorphism constraints
are invariant as
\bea
\delta{\cal H}^m= -{\cal H}^n\alpha_n{}^m~=~0~~~.
\eea
Noting  $\check{Z}_{mn}=\epsilon_{m n pq}\frac{1}{2}\epsilon^{ij}
\partial_i x^{p}
\partial_j x^{q} $ in \bref{eq4.1} it holds an identity
\bea
{\cal H}^\sharp\equiv \epsilon^{mnlp}\check{Z}_{mn}\check{Z}_{lp}=0~~.
\eea
It leads to the diffeomorphism constraints as a SL(5)  
vector form,
\bea
{\cal H}^{\hat{m}}=\frac{1}{8}\epsilon^{\hat{m}\hat{n}\hat{l}\hat{p}\hat{q}}
\check{Z}_{\hat{n}\hat{l}}\check{Z}_{\hat{p}\hat{q}}~~,
\hat{m}=(\sharp,~m)~.
\eea
Under the SL(5) transformation it is transformed as
\bea
&\check{Z}_{\hat{m}\hat{n}}~\to~g_{\hat{m}}{}^{\hat{m}'}
g_{\hat{n}}{}^{\hat{n}'}\check{Z}_{\hat{m}'\hat{n}'}~~,~~
g=1+A\in {\rm SL}(5)&\nn\\
 &\Rightarrow~
{\cal H}^{\hat{m}}~\to~
\epsilon^{\hat{m}\hat{n}\hat{l}\hat{p}\hat{q}}
g_{\hat{n}}{}^{\hat{n}'}
g_{\hat{l}}{}^{\hat{l}'}
g_{\hat{p}}{}^{\hat{p}'}
g_{\hat{q}}{}^{\hat{q}'}
\check{Z}_{\hat{n}'\hat{l}'}\check{Z}_{\hat{p}'\hat{q}'}
=
{\cal H}^{\hat{m}'}(g^{-1})_{\hat{m}'}{}^{\hat{m}}~~~.&
\eea

\vskip 5mm

It is mentioned that  O$(3,3)$ symmetry can be seen from the metric 
of the $\sigma^i$ -diffeomorphism constraint $\tilde{\rho}^{MN;p}$
in \bref{Hj}.
If we choose the direction of
 ${\cal H}^p$ vector to be ${\cal H}^1$ by 
 the SO($d$) rotation,
 then off-diagonal element $\delta^{1m}_{[nl]}$  becomes 
 $\delta_l^m$ for $l=2,\cdots,d$,
resulting  its rank to be  $(d-1)$.
Using an elementary matrix ${\cal P}$ it is written as
\bea
\tilde{\rho}^{MN;p}={\cal P}^T\left(
\begin{array}{cc|c}
0&{\bf 1}_{(d-1)}&0\\
{\bf 1}_{(d-1)}&0&0\\
\hline
0&0&0
\end{array}
\right){\cal P}~~~,\label{Od1d1}
\eea
which contain  O($d-1,d-1$) invariance manifestly.
This gauge choice may be related to the double field formalism.
The more general argument on the relation between 
M theory duality basis and double field theory basis
is given in \cite{Thompson:2011uw}.
 
\par\vskip 6mm
\vskip 6mm
\subsection{SL(5) duality transformation from M2 Hamiltonian }

It is known that there exists a global symmetry 
in $d$-dimensionally reduced  supergravity theory 
which is larger than the one for the pure gravity
theory, SL($d$,{\bf R})$\times$GL(1,{\bf R}).
For the supergravity theory with $T^4$ 
it is SL(5,{\bf R}) including the subgroup 
SL(4,{\bf R})$\times$GL(1,{\bf R}).

The gauge field $(\hat{C}_a)^M$ in \bref{Chat} is
recasted in the $\check{Z}_{\hat{m}\hat{n}}$ basis 
in \bref{tensor10} as
\bea
(\hat{C}_a)^MZ_M~=~\frac{1}{2}(\check{{C}}_a)^{\hat{m}\hat{n}}
Z_{\hat{m}\hat{n}}~~,~~
(\check{C}_a)^{\hat{m}\hat{n}}~=~
\left\{
\begin{array}{ccl}
(\check{C}_a)^{\sharp{n}}&=&
 e_a{}^n
\\
(\check{C}_a)^{{m}{n}}&=&
\frac{1}{2}\epsilon^{mnpq}(\hat{C}_a)_{pq}
\end{array}\right.~~,~~
\eea
which is also written as
\bea
(\check{C}_a)^{{m}{n}}=
\frac{1}{2}\epsilon^{npqm}e_a{}^lC_{pql}
=\tilde{C}^{[m}e_a{}^{n]}~~,~~\tilde{C}^m=\frac{1}{3!}\epsilon^{mnlp}C_{nlp}~~~.
\eea

The Hamiltonian constraint for a membrane in \bref{MM2} is written as 
\bea
{\cal H}_\perp~&=&~\frac{1}{2}Z_M{\cal M}^{MN}Z_N
~=~\frac{1}{8}\check{Z}_{\hat{m}\hat{n}}\check{{\cal M}}^{\hat{m}\hat{n};\hat{p}\hat{q}}
\check{Z}_{\hat{p}\hat{q}}\nn\\
\check{{\cal M}}^{\hat{m}\hat{n};\hat{p}\hat{q}}
&=&\frac{1}{4}W_{\hat{a}\hat{b}}{}^{\hat{m}\hat{n}}{}
\eta^{\hat{a}\hat{b};\hat{c}\hat{d}}
W_{\hat{c}\hat{d}}{}^{\hat{p}\hat{q}}{}~~,~~
\eta^{\hat{a}\hat{b};\hat{c}\hat{d}}=
\delta^{[\hat{a}|\hat{c}}
\delta^{|\hat{b}]\hat{d}}~,\nn~\label{geneM}\\
W_{\hat{a}\hat{b}}{}^{\hat{m}\hat{n}}&=&
\left(
\begin{array}{cc}
e_b{}^n& -\tilde{C}^{[m}e_b{}^{n]}\\
0& {\bf e}e_a{}^{[m} e_b{}^{n]}
\end{array}
\right)~~,~~
\label{WW}
\eea
with indices $(\hat{a}\hat{b})=(\natural b,~ab)$,
$(\hat{m}\hat{n})=(\sharp n, ~mn)$
 and ${\bf e}=\det e_m{}^a$.
 $W_{\hat{a}\hat{b}}{}^{\hat{m}\hat{n}}
$ is a
10$\times$10 matrix representation
of $\nu_{A}{}^{N}$  in \bref{munu}.
It is a coset element of 
SL(5)/SO(5) expressed by 
 $24-10=10+4=14$ coset parameters $G_{mn},~C_{mnl}$.
 The 10$\times$10 matrix
$W_{\hat{a}\hat{b}}{}^{\hat{m}\hat{n}}$ 
contains the $4\times 10$ matrix $(\check{C}_a)^{\hat{m}\hat{n}}$
manifestly and  the local SO(5) symmetry is used for
the triangular gauge.

Furthermore  $W$ in \bref{WW} is a tensor product of the $5\times 5$ representation
of  the coset SL(5)/SO(5). 
It is known that the ``generalized vielbein" for $T^4$ reduced supergravity 
as a coset element SL(5)/SO(5) 
derived in \cite{Cremmer:1979up,Sezgin:1982gi,Tanii:1984zk}
\bea
V_{\hat{m}}{}^{\hat{a}}~=~
\left(
\begin{array}{cc}
{\bf e}^{3/5}&{\bf e}^{-2/5}\tilde{C}^l e_l{}^a\\
0&{\bf e}^{-2/5}e_m{}^a
\end{array}
\right)~~,~~\det V=1~~.\label{VV}
\eea
The SL(5) invariant current is constructed as
$V^{-1}\partial_\mu V$.
Bilinear of its coset part gives the $d$-dimensional part of 
supergravity action.
It is transformed as $V~\to~gVh$ with $g\in$ SL(5) and $h\in$ SO(5).
After the SO(5) pull back,
the SL(5) transformation rules are given by
\bea
&&\delta V_{\hat{m}}{}^{\hat{a}}~=~
A_{\hat{m}}{}^{\hat{n}}V_{\hat{n}}{}^{\hat{a}}
+V_{\hat{m}}{}^{\hat{b}}
\lambda_{\hat{b}}{}^{\hat{a}}
\nn\\
&&~~~~\delta e_m{}^a~=~\alpha_m{}^ne_n{}^a
+\beta_m\tilde{C}^ne_n{}^a+e_m{}^b\lambda_b{}^{a}
-\frac{2}{3}(\hat{\alpha}+\tilde{C}^l\beta_l)e_m{}^a
\nn\\
&&{\renewcommand{\arraystretch}{1.2}
\left\{
\begin{array}{ccl}
\delta G_{mn}&=&\alpha_{(m|}{}^lG_{l|n)}
+\beta_{(m|}\tilde{C}^lG_{l|n)}
-\displaystyle\frac{4}{3}(\hat{\alpha}+\tilde{C}^l\beta_l)G_{mn}
\\
\delta \tilde{C}^m&=&-\hat{\alpha}\tilde{C}^m
-\tilde{C}^n\alpha_n{}^m
+{\bf e}^2\beta_nG^{nm}
-\beta_{l}\tilde{C}^l\tilde{C}^m+\gamma^m
\end{array}
\right.}\label{GCSL5}
\eea
where SO(4) symmetry parameter is $\lambda_{b}{}^a$.

The SL(5) tensor 
$W_{\hat{a}\hat{b}}{}^{\hat{m}\hat{n}}$ 
is rewritten by this ``generalized vielbein" as
\bea
W_{\hat{a}\hat{b}}{}^{\hat{m}\hat{n}}={\bf e}^{1/5}(V^{-1})_{\hat{a}}{}^{[\hat{m}}
(V^{-1})_{\hat{b}}{}^{\hat{n}]}~~,
~~
V^{-1}{}_{\hat{a}}{}^{\hat{m}}
=\left(
\begin{array}{cc}
{\bf e}^{-3/5}&-{\bf e}^{-3/5}\tilde{C}^m \\
0&{\bf e}^{2/5}e_a{}^m
\end{array}
\right)
\eea
then the ``generalized metric"  in \bref{geneM}
 is rewritten as
quartic in the generalized vielbein as
\bea
\check{\cal M}^{\hat{m}\hat{n};\hat{p}\hat{q}}&=&
\frac{{\bf e}^{2/5}}{4}
(V^{-1})_{\hat{a}}{}^{[\hat{m}}
(V^{-1})_{\hat{b}}{}^{\hat{n}]}
\eta^{\hat{a}\hat{b};\hat{a}'\hat{b}'}
(V^{-1})_{\hat{a}'}{}^{[\hat{p}}
(V^{-1})_{\hat{b}'}{}^{\hat{q}]}\nn\\
\nn\\
&=&
\left(
{\renewcommand{\arraystretch}{1.2}
\begin{array}{cc}
G^{nq}&-\tilde{C}^{[p}G^{q]n}\\
-\tilde{C}^{[m}G^{n]q}~~~&
{\bf e}^2G^{p[m}G^{n]q}+\tilde{C}^{[m}G^{n][q}\tilde{C}^{p]}
\end{array}}
\right)~,~(\hat{m}\hat{n})=(\sharp n,~mn).
\eea

The Hamiltonian constraint
 for a M2-brane in \bref{HamM2} is now written as
\bea
{\cal H}_\perp&=&\frac{{\bf e}^{2/5}}{8}
J_{\hat{a}\hat{b}}\eta^{\hat{a}\hat{b};\hat{a}'\hat{b}'}
J_{\hat{a}'\hat{b}'}~~,~~
J_{\hat{a}\hat{b}}~=~
\frac{1}{2}(V^{-1})_{\hat{a}}{}^{[\hat{m}}\check{Z}_{\hat{m}\hat{n}}
(V^{-1}{}^T)^{\hat{n}]}{}_{\hat{b}}
\eea
where the current $J_{\hat{a}\hat{b}}$ is manifestly SL(5)
invariant and SO(5) invariant 
to guarantee the SL(5) covariance of the system.
By choosing a gauge of the $\tau$-diffeomorphism
invariance ${\bf h}={\bf e}^{-2/5}$ 
\bea
 H=\displaystyle\int d^2\sigma~{\bf h} {\cal H}_\perp
 ~,~{\bf h}={\bf e}^{-2/5} ~\Rightarrow ~
 H=\displaystyle\int d^2\sigma~\frac{1}{8}
J_{\hat{a}\hat{b}}\eta^{\hat{a}\hat{b};\hat{a}'\hat{b}'}
J_{\hat{a}'\hat{b}'}~~,
 \eea
then the Hamiltonian has manifest SL(5) invariance.
\footnote{
This is related to the fact that dilaton field is required to construct
the Ricci scalar from T-dual manifest covariant derivatives as shown in
\cite{Siegel:1993th}. We thank Warren Siegel for explaining this. It is 
also mentioned that the enlargement of dimensions may be related to F-theory.
}
The SL(5) duality transformation of $G_{mn}$ and $C_{mnl}$
are given in \bref{GCSL5}.

\vskip 6mm
\section{Summary and discussion}

We have seen how the SL(5) duality in M-theory 
is derived from the M2-brane mechanics.
A Courant bracket for a M2-brane is obtained 
to derive the generalized gauge transformation rules
for $G_{mn}$ and $C_{mnl}$.
 It is natural to use the anti-symmetric rank two basis $\check{Z}_{\hat{m}\hat{n}}$ as 10-representation of SL(5). 
In this  basis 
 the diffeomorphism constraints are rewritten
as a SL(5) vector showing the SL(5) invariance 
on the constraint surface.
The Hamiltonian is written by 
the rank four generalized metric
which is quartic in the generalized vielbein. 
The generalized vielbein is an element of the coset  SL(5)/SO(5)
parameterized by $G_{mn}$ and $C_{mnl}$.  
Thus the SL(5) duality  transformation of $G_{mn}$
and $C_{mnl}$ is obtained as that of the coset parameter.
 The diffeomorphism and Hamiltonian constraints 
 in a background dependent gauge are invariant under the 
SL(5) transformations of the basis $\check{Z}_{\hat{m}\hat{n}}
\to g_{\hat{m}}{}^{\hat{m}'}g_{\hat{n}}{}^{\hat{n}'} Z_{\hat{m}'\hat{n}'}$, 
when the  coset parameters 
$G_{mn},~C_{mnl}$ are transformed in the non-linear realization of SL(5)/SO(5).

The extension of this analysis to M5-brane system 
involves  16-representation of SO(5,5) duality symmetry in $d=5$.
The 16-representation of SO(5,5) contains $5+10+1$  
which corresponds to the usual vector,
2-form and $5$-form in $d=5$.   
Including such higher dimensional cases manifestation of
 duality symmetries is forthcoming problem.
Treatment of RR gauge fields and fermionic fields
are also necessary to be clarified.

\vskip 6mm
\section*{Acknowledgements}
M.H. would like to thank Warren Siegel and Maxim Zabzine  for fruitful discussions.
She is also  grateful to the 2012 Summer Simons workshop in Mathematics and Physics
 for a stimulating environment and its warm hospitality. 
 The work of M.H. is supported  by Grant-in-Aid for Scientific Research (C) No. 24540284 from The Ministry of Education, Culture, Sports, Science and Technology of Japan.

\end{document}